\documentclass[%
 reprint,
nobibnotes,
 amsmath,amssymb,
 aps,
prl,
twocolumn,
]{revtex4-1}

\usepackage{graphicx}
\usepackage{dcolumn}
\usepackage{bm}
\usepackage{enumerate}
\usepackage{verbatim}
\usepackage{float}
\usepackage{here}

\usepackage{color}
\usepackage[bookmarks,colorlinks,linkcolor=webblue,citecolor=webgreen]{hyperref}	
\definecolor{webgreen}{rgb}{0, 0.5, 0} 
\definecolor{webblue}{rgb}{0, 0, 0.5} 
\definecolor{webred}{rgb}{0.5, 0, 0} 

\newcommand{\di}{\text{d}}

\begin{document}

\preprint{}
\title{Simultaneously solving the $H_0$ and $\sigma_8$ tensions with late dark energy}
\author{Lavinia Heisenberg\footnote{lavinia.heisenberg@phys.ethz.ch}, 
        Hector Villarrubia-Rojo\footnote{herojo@phys.ethz.ch},
        Jann Zosso\footnote{jzosso@phys.ethz.ch}}
\affiliation{Institute for Theoretical Physics, ETH Z\"{u}rich, Wolfgang-Pauli-Strasse 27, 8093, Z\"{u}rich, Switzerland}

\date{\today}

\begin{abstract}
In a model independent approach, we derive generic conditions that any late time modification 
of the $\Lambda$CDM expansion history must satisfy in order to consistently solve both the 
$H_0$ and the $\sigma_8$ tensions. Our results are fully analytical and the method is 
merely based on the assumption that the late-time deviations from $\Lambda$CDM remain 
small. For the concrete case of a dark energy fluid with deviations encoded in the expansion 
history and the gravitational coupling constant, we present necessary conditions on
its equation of state. Solving both the $H_0$ and $\sigma_8$ tensions requires that $w(z)$ 
must cross the phantom divide if $G_\text{eff}=G$. On the other hand, for $G_\text{eff}=G+\delta G(z)$ and $w(z)\leq -1$,
it is required that $\displaystyle \frac{\delta G(z)}{G}<\alpha(z)\frac{\delta H(z)}{H(z)}<0$ at some redshift $z$.
\end{abstract}

\maketitle


\section{Introduction}\label{sec:introduction}

Todays era of precision cosmology poses a serious challenge to the theoretical description 
of our universe providing a guideline on the search for new physics. While the dark sector 
of the current cosmological standard model, that is a cosmological constant $\Lambda$ 
together with cold dark matter ($\Lambda$CDM), remains poorly understood, the data is 
beginning to require a departure from the model itself with increasing statistical 
significance. This is mainly due to the mismatch or tensions between the values of 
cosmological observables inferred from the cosmic microwave background (CMB) on the one 
hand and several independent local measurements on the other. 
The two most prominent tensions are discrepancies in the values of the Hubble 
constant $H_0$ and the clustering amplitude $\sigma_8$ arising between the Planck values 
\cite{Planck:2018vyg} in comparison to long-period Cepheid, Megamaser and strong lensing 
observations \cite{Riess:2019cxk,Riess:2020fzl,Pesce:2020xfe,Wong:2019kwg} and large scale structure 
(LSS) surveys \cite{DES:2017myr, DES:2021wwk, KiDS:2020suj, Heymans:2020gsg} respectively. Concretely, the value of $H_0$ 
as inferred by Planck using $\Lambda$CDM is between $4.5\sigma$ to $6.3\sigma$ below late 
time measurements \cite{Riess:2019qba} depending on different measurement combinations, 
whereas the Plank value of $\sigma_8$ (or $S_8$) is systematically above low redshift estimates 
\cite{Heymans:2020gsg,Nunes:2021ipq}, although at a lower statistical significance. 

While the multitude of independent measurements progressively supporting the tensions 
demand for a theory beyond $\Lambda$CDM, there is no lack of alternatives. An abounding 
amount of models have been proposed as a solution to one or both of the tensions (see for instance
\cite{Knox:2019rjx,DiValentino:2021izs,DiValentino:2020zio,DiValentino:2020vvd, Perivolaropoulos:2021jda, Poulin:2018cxd, Smith:2019ihp, Alcaniz:2019kah, Zumalacarregui:2020cjh, Gomez-Valent:2020mqn, Ballesteros:2020sik, Jimenez:2020bgw, DiValentino:2020naf, Lambiase:2018ows, 
Keeley:2019esp, DiValentino:2019ffd, Jedamzik:2020zmd, Clark:2021hlo, SolaPeracaula:2021gxi,
Alestas:2021xes, Nunes:2021ipq, Schoneberg:2021qvd, Alestas:2021luu, Renk:2017rzu,Frusciante:2019puu,deFelice:2017paw,DeFelice:2020sdq,Heisenberg:2020xak}). Fortunately, the growing web of precise cosmological surveys pose tight 
constraints on any alternative model which is a reason to hope that cosmology will guide us 
towards the next step in fundamental physics.

Indeed, in this letter we present a model independent approach which, based on the precise 
value of the CMB acoustic scale and a minimal set of assumptions alone, allows us to derive 
necessary conditions which severely constrain late time dark energy (DE) models 
characterized by an equation of state $w(z)$ which deviate from $\Lambda$CDM through the 
expansion history $\delta H(z)$ and a modification of the gravitational constant 
$\delta G(z)$. For any such theory we show the following:
\newline

\noindent\emph{Solving both the $H_0$ and $\sigma_8$ tensions requires:}
\begin{enumerate}[i)]
\item \emph{If $G_\text{eff}=G$, $w(z)$ must cross the phantom divide.}
\item \emph{If $G_\text{eff}=G+\delta G(z)$ and $w(z)\leq -1$}, \\[5pt]
	$\displaystyle \frac{\delta G(z)}{G}<\alpha(z)\frac{\delta H(z)}{H(z)}<0\;\;\text{\emph{at some redshift}}\;z$,
\end{enumerate}
where in the second case, due to the condition on the equation of state, $\delta H(z)\neq 0$
and $\alpha(z)$, which we define in \eqref{eq:alpha}, is strictly positive.

In the following we will describe the method which allows us to arrive at the above results, 
referring to the companion paper \cite{companionpaperHRZ} for details. \emph{We want to stress 
that our method is applicable in a much broader context and it is a priori not tied to dark 
energy models nor any specific cosmological tension.}


\section{Method}\label{sec:method}

\hspace*{1mm}\noindent\textit{Setup.}\hspace*{2mm}
The starting point is a $\Lambda$CDM cosmology, which at late times can effectively be 
described by two free parameters, the Hubble constant $H_0$ and the matter abundance 
$\Omega_m$ through
\begin{equation}
	H_\text{\tiny $\Lambda$CDM}^2= H_0^2 \left(\Omega_m (1+z)^{3}+ \Omega_\Lambda\right)\,.
\end{equation}
In the following, we will define $\omega_m=\Omega_m h^2$ and $\omega_\Lambda=\Omega_\Lambda h^2$, where 
$H_0\equiv 100 \ h \ \text{km}\,\text{s}^{-1}\,\text{Mpc}^{-1}$, such that 
\begin{equation}
	\omega_\Lambda=h^2-\omega_m\,.
\end{equation}

Alternative models can then be characterized by variations of the expansion history 
$\delta H(z)$ 
\begin{equation}
	H(H_0, \omega_m) = H_\text{\tiny $\Lambda$CDM}(H_0,\omega_m) + \delta H(z)\ ,
\end{equation}
as well as variations in other quantities, such as the gravitational constant 
$G_\text{eff}=G+\delta G(z)$. For 
brevity, however, we will focus on $\delta H(z)$ during the exposition of the method and 
refer to the last paragraph of this section for the general case. At this stage, 
$\delta H(z)$ is an arbitrary function which captures deviations from $\Lambda$CDM for 
fixed $H_0$ and $\omega_m$. Restricting ourselves to late-time modifications, we will
assume that $\delta H(z) = 0$ for $z>300$. 

The deviation from $\Lambda$CDM will modify the observationally preferred values of $H_0$ and $\omega_m$ such that working at first order in deviations the Hubble parameter in the alternative cosmology takes the general form
\begin{equation}
	H(H_0+\delta H_0, \omega_m+\delta\omega_m) = H_\text{\tiny $\Lambda$CDM}(H_0,\omega_m) + \Delta H\ .
\end{equation}
For late-time modifications, the deviations in the matter abundances are generally 
negligible, see \cite{companionpaperHRZ}, so for simplicity we impose $\delta\omega_m=0$ 
in the reminder of this letter. In this case, the total variation in the Hubble parameter 
reads
\begin{equation}\label{eq:DeltaH}
	\frac{\Delta H(z)}{H(z)} = \frac{H_0^2}{H^2(z)}\frac{\delta H_0}{H_0} +  \frac{\delta H(z)}{H(z)}\ ,
\end{equation}
where, since we are working to first order, we are denoting $H_\text{\tiny $\Lambda$CDM}$
simply as $H$.
This allows us to express the variation of any cosmological observable $\mathcal{O}$ as
\begin{equation}\label{eq:VarGen}
	\frac{\Delta\mathcal{O}(z)}{\mathcal{O}(z)} = I_\mathcal{O}(z)\frac{\delta H_0}{H_0}+ \int^\infty_0\frac{\di x_z}{1+x_z}R_\mathcal{O}(x_z, z)\frac{\delta H(x_z)}{H(x_z)}\ .
\end{equation}

\hspace*{1mm}\noindent\textit{Relation to observations.}\hspace*{2mm}
It is now enough to choose a single very well measured observable, $\mathcal{O}_*$, whose 
value should not change in the alternative model with the aim of remaining compatible with 
observations, and hence impose its variation to vanish, $\Delta \mathcal{O}_*=0$, in order 
to relate the modified expansion history $\delta H(z)$ to the variation in the inferred 
Hubble constant $\delta H_0$ through a response function
\begin{align}\label{eq:Resph}
	\frac{\delta H_0}{H_0} &= -\int\frac{\di x_z}{1+x_z}\frac{R_{\mathcal{O}_*}(x_z)}{I_{\mathcal{O}_*}}\frac{\delta H(x_z)}{H(x_z)}\nonumber\\
						   &\equiv \int\frac{\di x_z}{1+x_z}\mathcal{R}_{H_0}(x_z)\frac{\delta H(x_z)}{H(x_z)}\,.
\end{align}
Of course, agreeing with just one observable is not enough for a model to be viable. 
However, this simple method will allow us to derive conditions that a model must
\emph{at least} satisfy in order not to be directly excluded. And these necessary conditions will pose stringent analytic constraints on the allowed modifications in the expansion history.

\hspace*{1mm}\noindent\textit{Deriving necessary conditions.}\hspace*{2mm}
Combining \eqref{eq:VarGen} and \eqref{eq:Resph} allows to compute the response function of 
arbitrary observables
\begin{equation}\label{eq:RespGen}
	\frac{\Delta\mathcal{O}(z)}{\mathcal{O}(z)} =\int^\infty_0\frac{\di x_z}{1+x_z}\, \mathcal{R}_\mathcal{O}(x_z, z)\frac{\delta H(x_z)}{H(x_z)}\ .
\end{equation}
It is now possible to formulate necessary conditions on the functional form of $\delta H$ 
in order to achieve the desired modifications in the inferred values of the quantities
that could alleviate the tensions, e.g. decreasing $\sigma_8$. These conditions crucially 
depend on the shape of the response functions.

\hspace*{1mm}\noindent\textit{Generalizations.}\hspace*{2mm}
As shown in the companion paper \cite{companionpaperHRZ}, various generalizations to the 
above method are possible. First of all, everything goes through without imposing 
$\delta\omega_m=0$ which is only a valid approximation for low redshift modifications of 
$\Lambda$CDM. In fact, the method is a priori not even tight to choosing $\Lambda$CDM as a 
starting point such that deformations of generic Hubble parameter functions can be considered. 
Moreover, as already mentioned, it is possible to allow for deviations of additional 
quantities $\delta Q_i(z)$ in form of generic functions. Typically, alternative theories to 
$\Lambda$CDM involve such additional deviations at the level of perturbations. Considering 
all this, \eqref{eq:VarGen} generalizes to
\begin{align}\label{eq:VarAll}
	\frac{\Delta\mathcal{O}}{\mathcal{O}} = I_\mathcal{O}\frac{\delta H_0}{H_0}&+ J_\mathcal{O}\frac{\delta\omega_m}{\omega_m}+ \int^\infty_0\frac{\di x_z}{1+x_z}R_\mathcal{O}(x_z)\frac{\delta H(x_z)}{H(x_z)}\nonumber\\
		&+\sum_i \int^\infty_0\frac{\di x_z}{1+x_z}\mathcal{Q}_{i\mathcal{O}}(x_z)\frac{\delta Q_i(x_z)}{Q_i(x_z)}.
\end{align}
As generality increases, however, constraining $\delta H$ as well as $\delta Q_i$ will 
require additional conditions such as imposing multiple observational constraints.
	

\section{Results}\label{sec:results}

We will now present our main results, which follow from applying the previous method to the 
specific context of the $H_0$ and $\sigma_8$ tensions and derive general requirements on the 
functional form of deviations from $\Lambda$CDM in order to achieve desired variations in 
the Hubble constant and the clustering amplitude. For computations and various details we 
refer the reader to the companion paper \cite{companionpaperHRZ}.

The first task is to obtain an analytic expression for $\sigma_8$, the amplitude of the 
density fluctuation power spectrum evaluated on spheres of radius $R=8\,h^{-1}$ Mpc \footnote{See e.g. \cite{dodelson2020modern} for a definition of $\sigma_R$.}, in order to compute its variation in the form of \eqref{eq:VarAll}. This is possible by adopting the Eisenstein-Hu fitting formula \cite{Eisenstein:1997ik} that takes into account the baryonic suppression at small scales which proves important for an accurate computation of the clustering amplitude. The resulting expression is proportional to the growing mode of the growth factor which for sub horizon modes and neglecting radiation as well as neutrino masses can be expressed analytically as well. See \cite{companionpaperHRZ} for the final expression.

\subsection{Modifying the expansion history}\label{sec:dH}
A broad class of proposed solutions to the $H_0$ tension modify the $\Lambda$CDM 
background without introducing significant deviations in the perturbations, i.e.
without introducing new clustering species or modifying quantities like the 
gravitational coupling $G_\text{eff}\simeq G$.

\begin{figure}[H]
\centering
\includegraphics[scale=0.42]{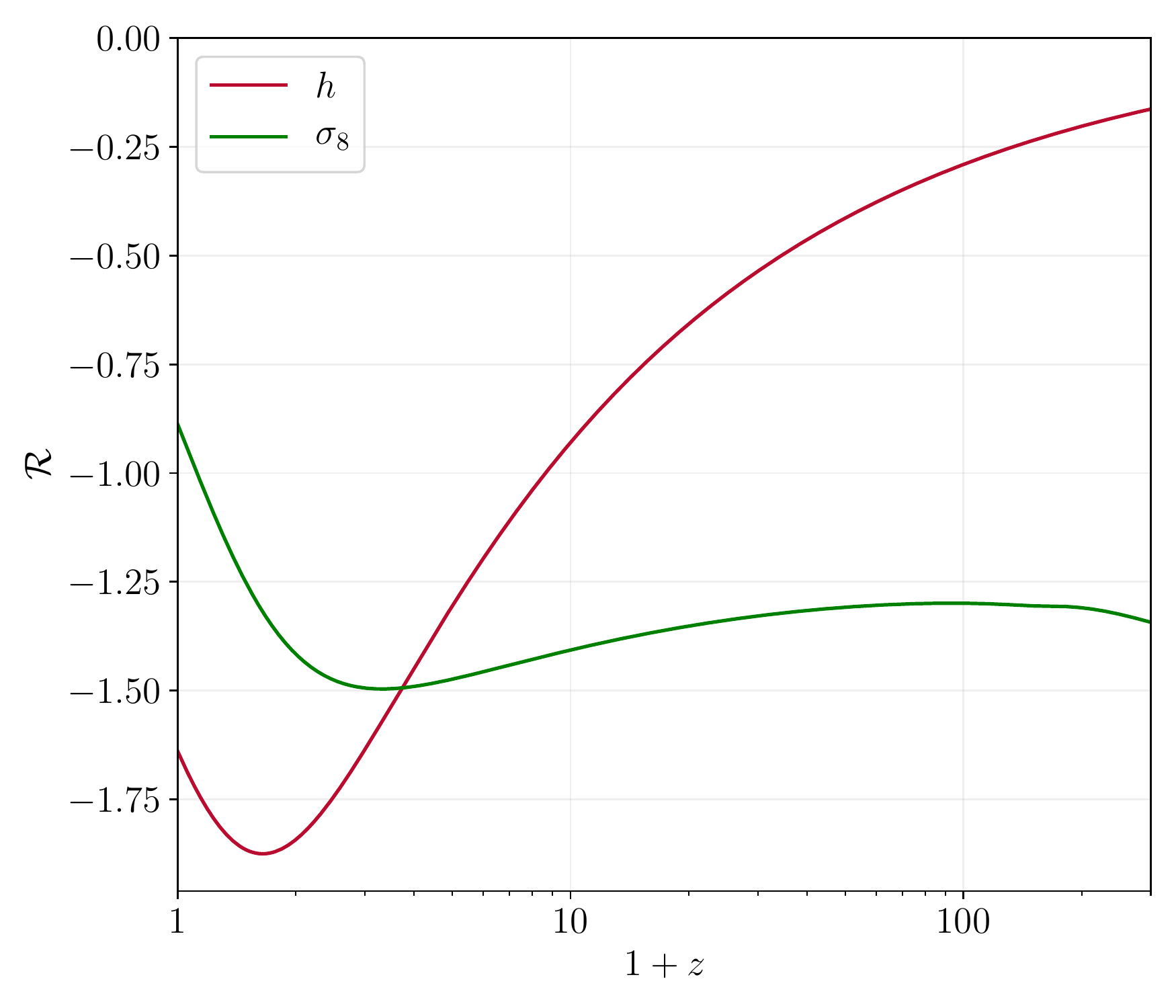}
\caption{\label{fig:responses}The response functions $\mathcal{R}_{H_0}(z)$ and $\mathcal{R}_{\sigma_8}(z, 0)$ as defined in \eqref{eq:Resph} and \eqref{eq:RespGen} respectively. Both responses remain strictly negative over the entire rage $0<z<300$ in which the expansion history is modified.}
\end{figure}

In these models, the variation of $\sigma_8$ takes the form of \eqref{eq:VarGen}. As discussed above, working with fixed 
$\omega_m$ without loss of generality as confirmed in \cite{companionpaperHRZ}, we now 
merely need to impose a zero variation of a single observable in order to relate the 
deviation $\delta H(z)$ from $\Lambda$CDM to $\delta H_0$ and subsequently to 
$\Delta\sigma_8$. In this work we will choose the variation of the CMB acoustic scale 
$\theta_*$ \cite{Chen:2018dbv} to vanish. The resulting response function 
$\mathcal{R}_{H_0}(z)$ as defined in \eqref{eq:Resph} is used to calculate the response 
function of $\sigma_8$ today as
\begin{equation}
	\mathcal{R}_{\sigma_8}(z, 0) = I_{\sigma_8}(0)\mathcal{R}_{H_0}(z) + R_{\sigma_8}(z, 0)\,.
\end{equation}
The results are plotted in Fig.~\ref{fig:responses}. The shape of the response functions 
immediately allows to draw simple conclusions with substantial impact. From 
\eqref{eq:Resph} alone, it follows that in order solving the $H_0$ tension, hence a 
positive $\delta H_0$, necessarily requires that 
\begin{equation}
	\delta H_0>0\ \implies\ \delta H(z)<0\ \text{at some }z\,.
\end{equation}
On the other hand, relieving the $\sigma_8$ tension, which requires a negative variation 
$\Delta\sigma_8$, is only possible if
\begin{equation}
	\Delta\sigma_8<0\ \implies\ \delta H(z)>0\ \text{at some }z\,.
\end{equation}
In other words, increasing the value of the Hubble constant while simultaneously 
decreasing the clustering amplitude necessarily requires $\delta H(z)$ to change sign.

This general result can readily be used to rule out specific models proposed in the 
literature. For example, for late dark energy with equation of state $w(z)$ and under the 
assumption that DE perturbations only play a subleading role in driving the values of 
$H_0$ and $\sigma_8$, which is the case for many typical theories, we conclude that for 
such models \emph{increasing $H_0$ while decreasing $\sigma_8$ necessarily requires $w(z)$ 
to cross the value $w=-1$}, since the sign of $\delta H(z)$ is directly connected to the 
sign of $1+w(z)$ \cite{companionpaperHRZ}.

\subsection{Beyond the expansion history: $G_\text{eff}$}\label{sec:dG}

\begin{figure}[H]
\centering
\includegraphics[scale=0.4]{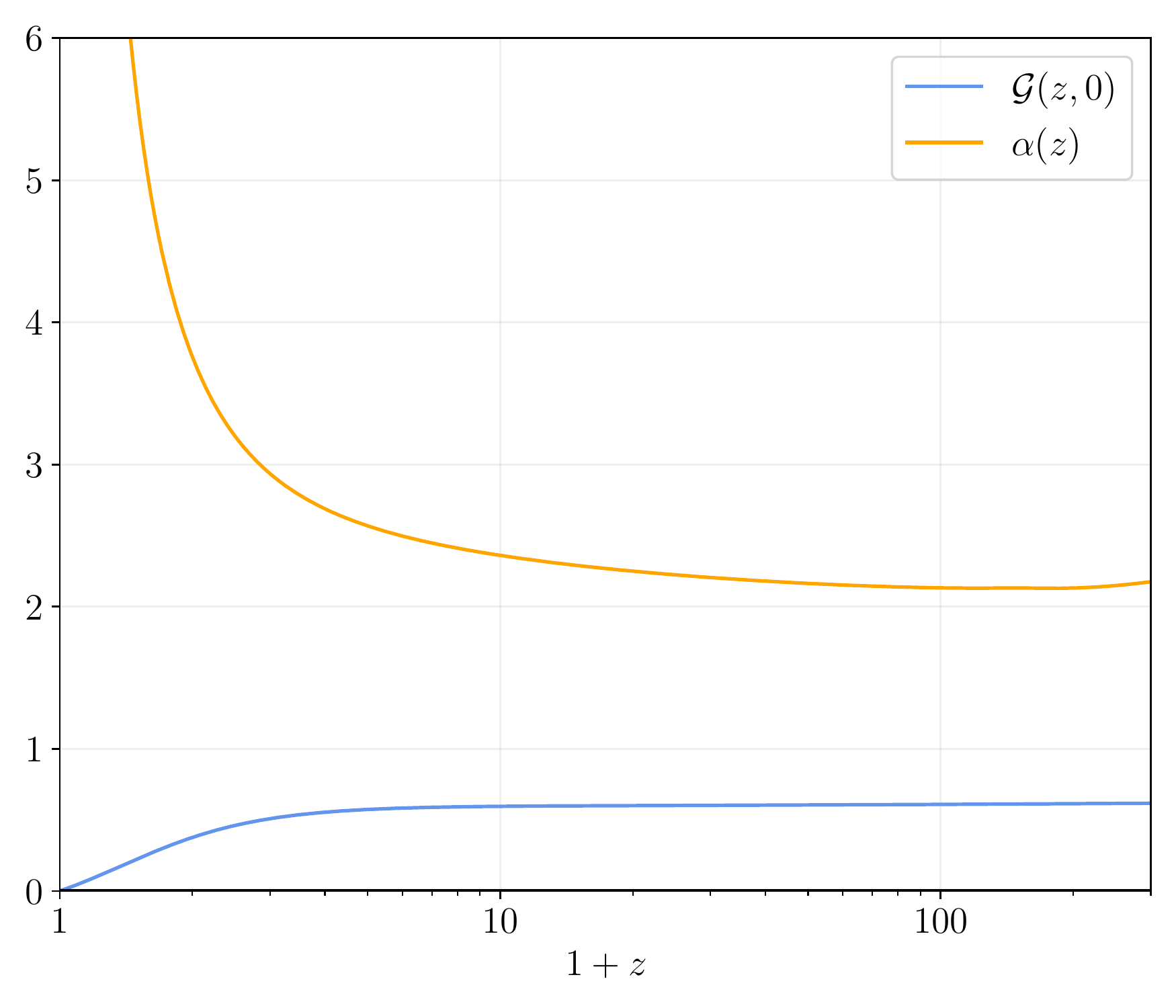}
\caption{\label{fig:responsesG}The response function $\mathcal{G}_{\sigma_8}(z, 0)$ and the function $\alpha(z)$ as defined in \eqref{eq:RespGen} and \eqref{eq:alpha} respectively. Both functions remain strictly positive over the entire rage $0<z<300$ in which the expansion history is modified.}
\end{figure}

As remarked in the previous section, going beyond the background evolution induces 
additional small deviations from $\Lambda$CDM, parametrized by additional functions 
$\delta Q_i(z)$. For example, dark energy clustering or modified gravity typically 
introduce modifications to the gravitational coupling $G_\text{eff}=G+\delta G$. To first 
order, this will only affect the growth factor and therefore the variation of the 
clustering amplitude via
\begin{equation}\label{eq:RespGen}
\frac{\Delta\sigma_8}{\sigma_8} =\int^\infty_0\frac{\di x_z}{1+x_z}\, \mathcal{R}_{\sigma_8}\frac{\delta H}{H}+\int^\infty_0\frac{\di x_z}{1+x_z}\, \mathcal{G}_{\sigma_8}\frac{\delta G}{G}\,.
\end{equation}
The result for $\mathcal{G}_{\sigma_8}(z,0)$ derived in \cite{companionpaperHRZ} is again 
presented visually in Fig.~\ref{fig:responsesG}. Given our minimal approach, that is 
only enforcing one observational anchor point given by the CMB acoustic scale, it is not 
possible to derive similarly strong conditions on the two free functions $\delta H$ and 
$\delta G$ independent of any quantitative analysis. However, if we restrict ourselves to 
cases where $\delta H<0$ (i.e. $w(z)\leq-1$), the following must hold
\begin{align}
\Delta\sigma_8<0\ \implies\ \frac{\delta G(z)}{G}<\alpha(z)\frac{\delta H(z)}{H(z)}<0\ \text{at some } z\,,
\end{align}
where we have defined the strictly positive function
\begin{align}\label{eq:alpha}
\alpha(z)\equiv - \frac{\mathcal{R}_{\sigma_8}(z,0)}{\mathcal{G}_{\sigma_8}(z,0)}\,.
\end{align}

\section{Discussion}\label{sec:discussion}

The methodology developed in this work has allowed us to identify a class of models
that can \emph{not} solve both the $H_0$ and $\sigma_8$ tensions. Focusing on late DE models, with equation of state $w(z)$ and with a modified gravitational
coupling $G_\text{eff}=G+\delta G(z)$, we derived the necessary conditions that must be met to 
solve the $H_0$ and the $\sigma_8$ tensions, i.e. $\delta H_0>0$ and $\Delta\sigma_8<0$, 

\begin{enumerate}[i)]
	\item Solving the $H_0$ tension $\implies$ $w(z)<-1$ at some $z$.
	\item If $G_\text{eff}=G$:\\[3pt]
		Solving the $H_0$ and $\sigma_8$ tensions $\implies$\\[3pt]
		$w(z)$ must cross phantom limit $w=-1$ at some $z$.
	\item If $G_\text{eff}=G+\delta G$ and $w(z)\leq-1$ (i.e. $\delta H<0$):\\[3pt]
		Solving the $H_0$ and $\sigma_8$ tensions $\implies$\\[3pt]
		$\displaystyle \frac{\delta G}{G}<\alpha(z)\frac{\delta H}{H}<0$ at some $z$, where $\alpha(z)>0$.		
\end{enumerate}

We would like to stress that despite the generality of these conditions obtained under 
a minimal set of assumptions, their implications are already significant. In words, a 
typical late dark energy model trying to ease the Hubble tension through a phantom equation 
of state will at the same time worsen the $\sigma_8$ tension if it does not either cross 
the phantom divide or significantly decrease $G_\text{eff}$. However, what is predominately 
observed is that late DE models increase the effective gravitational constant though clustering, 
while crossing the $w=-1$ surface is not trivial at all, potentially leading to divergences 
in dark energy perturbations.
 
Including quantitative arguments and additional observational constraints, it actually 
seems likely that any reasonably simple late DE model focusing on background evolution only 
will eventually fall short. Indeed, preliminary analysis in \cite{companionpaperHRZ} shows 
that models with changing sign of $\delta H(z)$ are not able to shift $H_0$ enough to fully 
resolve the tensions. On the other hand, including perturbations often worsens the situation.

It should be noted at this point that so-called early time solutions to the Hubble tension 
are by no means in a better position. For example, it was recently shown that early time 
solutions which solely reduce the cosmic sound horizon generally fall short  
\cite{Jedamzik:2020zmd}.

In this context, the analytic method developed in this work, starting with the presented results, is able to give valuable insights into the behavior of the dark sector, possibly providing hints towards building successful models beyond $\Lambda$CDM.

\section*{Acknowledgements}
We would like to thank Adam Riess for useful discussions and going through the draft.
LH is supported by funding from the European Research Council (ERC) under the European Unions Horizon 2020 research and innovation programme grant agreement No 801781 and by the Swiss National Science Foundation grant 179740. 

\bibliography{Biblio.bib}
\end{document}